# Pressure-induced Topological and Structural Phase Transitions in an Antiferromagnetic Topological Insulator*

Cui-Ying Pei(裴翠颖)[1,***], Yun-You-You Xia(夏云悠悠)[1,2,3,***], Jia-Zhen Wu(邬家臻)[4], Yi Zhao(赵毅)[1], Ling-Ling Gao(高玲玲)[1], Tian-Ping Ying(应天平)[4], Bo Gao(高波)[5], Na-Na Li(李娜娜)[5], Wen-Ge Yang(杨文革)[5], Dong-Zhou Zhang(张东舟)[6], Hui-Yang Gou(猴慧阳)[5], Yu-Lin Chen(陈宇林)[1,7,8], Hideo Hosono(细野秀雄)[4], Gang Li(李刚)[1,8,**] and Yanpeng Qi(齐彦鹏)[1,**]

[1] School of Physical Science and Technology, ShanghaiTech University, 393 Middle Huaxia Road, Shanghai 201210, China
[2] Shanghai Institute of Optics and Fine Mechanics, Chinese Academy of Sciences, Shanghai 201800, China
[3] University of Chinese Academy of Sciences, Beijing 100049, China
[4] Materials Research Center for Element Strategy, Tokyo Institute of Technology, 4259 Nagatsuta, Midori-ku, Yokohama 226-8503, Japan
[5] Center for High Pressure Science and Technology Advanced Research (HPSTAR), 1690 Cailun Road, Shanghai 201203, China
[6] Hawai'i Institute of Geophysics and Planetology, School of Ocean and Earth Science and Technology, University of Hawai'i at Manoa, Honolulu, Hawaii 96822 USA
[7] Department of Physics, Clarendon Laboratory, University of Oxford, Parks Road, Oxford OX1 3PU, UK
[8] ShanghaiTech Laboratory for Topological Physics, ShanghaiTech University, Shanghai 200031, China

**Corresponding author. Email: qiyp@shanghaitech.edu.cn; ligang@shanghaitech.edu.cn
*** These authors contributed equally.

Recently, natural van der Waals heterostructures of $(MnBi_2Te_4)_m(Bi_2Te_3)_n$ have been theoretically predicted and experimentally shown to host tunable magnetic properties and topologically nontrivial surface states. In this work, we systematically investigate both the structural and electronic responses of $MnBi_2Te_4$ and $MnBi_4Te_7$ to external pressure. In addition to the suppression of antiferromagnetic order, $MnBi_2Te_4$ is found to undergo a metal-semiconductor-metal transition upon compression. The resistivity of $MnBi_4Te_7$ changes dramatically under high pressure and a non-monotonic evolution of $\rho(T)$ is observed. The nontrivial topology is proved to persists before the structural phase transition observed in the high-pressure regime. We find that the bulk and surface states respond differently to pressure, which is consistent with the non-monotonic change of the resistivity. Interestingly, a pressure-induced amorphous state is observed in $MnBi_2Te_4$, while two high pressure phase transitions are revealed in $MnBi_4Te_7$. Our combined theoretical and experimental research establishes $MnBi_2Te_4$ and $MnBi_4Te_7$ as highly tunable magnetic topological insulators, in which phase transitions and new ground states emerge upon compression.

**PACS:** 03.65.Vf , 64.70.Tg, 07.35.+k,

Magnetic topological insulator (MTI), possessing both magnetic and topological properties, provide a promising material platform for the realization of exotic topological quantum phenomena, such as the quantum anomalous Hall (QAH) effect, axion insulator states, the proximity effect, Majorana modes, etc.[1-6] Thereinto the QAH effect has been observed experimentally in magnetically doped topological insulator (TI) thin films,[7] while the fabrication of homogeneous thin films has long been limited by deposition techniques, hindering extensive studies of the unique material systems. Hence intrinsic MTI with homogeneous magnetic and electronic properties is desired and can provide new opportunities to study novel topological quantum phenomena.

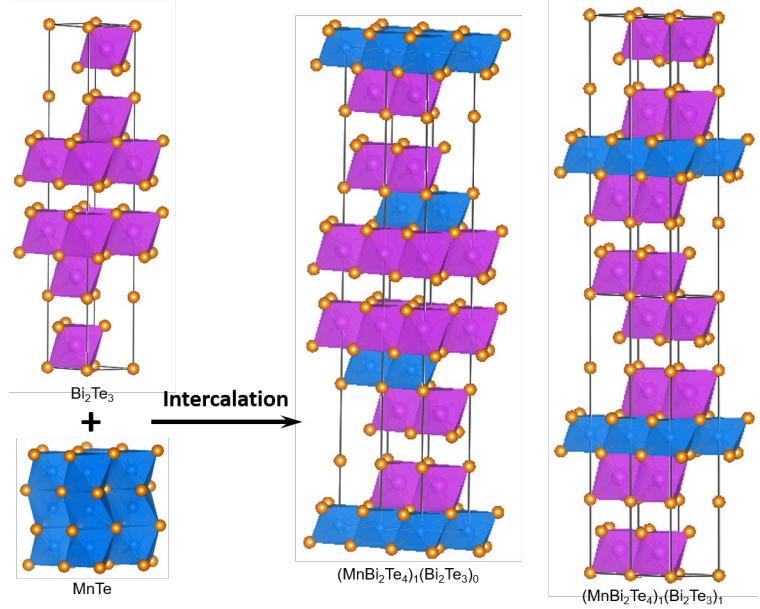

Fig. 1. Crystal structure of $Bi_2Te_3$ (*R*-3*m*, no. 166)[8], MnTe (*P*6$_3$/*mmc*, no. 194)[9], $MnBi_2Te_4$ (($MnBi_2Te_4$)$_1$($Bi_2Te_3$)$_0$, *R*-3*m*, no. 166)[10] and $MnBi_4Te_7$ (($MnBi_2Te_4$)$_1$($Bi_2Te_3$)$_1$, *P*-3*m*1, no. 164)[11], respectively. The $MnBi_2Te_4$ unit cell consists of three septuplet monoatomic layers with a stacking sequence of Te(1)-Bi(1)-Te(2)-Mn(1)-Te(2)-Bi(1)-Te(1) along the *c*-axis, and the seven monoatomic layers are centro-symmetrical with respect to Mn. In detail, Mn crystallographic sites are of octahedral coordination and are surrounded by six Te(2) atoms at the same distance as under ambient conditions. Bi is at the center of a distorted octahedron and is surrounded by three Te(2) atoms and three Te(1) atoms as the nearest neighbors. Triple slabs of $MnTe_6$ and $BiTe_6$ are octahedral edge-linked with each other, and similarly for the SL of $MnBi_4Te_7$. Alternation of QL (Te(2)-Bi(1)-Te(1)-Bi(1)-Te(2)) and SL (Te(3)-Bi(2)-Te(4)-Mn(1)-Te(4)-Bi(2)-Te(3)) blocks stack along the *c*-axes and $MnBi_4Te_7$ is in the trigonal space group *P*-3*m*1.

Recently intrinsic MTIs of $(MnBi_2Te_4)_m(Bi_2Te_3)_n$ has been theoretically predicted and experimentally synthesized to have tunable magnetic properties and topologically nontrivial surface states.[11-23] As shown in Figure 1, $(MnBi_2Te_4)_m(Bi_2Te_3)_n$ crystallizes in a van der Waals layered structure, sharing a similar crystal structure with $Bi_2Te_3$[8], a typical TI in ambient conditions. Crystalizing in a rhombohedral structure with space group *R*-3*m*, $MnBi_2Te_4$ ($m = 1$, $n = 0$) consists of Te-Bi-Te-Mn-Te-Bi-Te septuple layers (SLs) as the building block, each of which can be viewed as a $Bi_2Te_3$ quintuple layer (QL) intercalated by a MnTe bilayer. $MnBi_4Te_7$ ($m = 1$, $n = 1$) adopts space group *P*-3*m*1 with a hexagonal superlattice crystal structure with alternate stacking of one $MnBi_2Te_4$ SL and one $Bi_2Te_3$ QL. $MnBi_2Te_4$ and $MnBi_4Te_7$ are both identified to be natural van der Waals heterostructures as evidenced by high-angle annular dark field (HAADF)-STEM measurements.[12]

Pressure as a conventional thermodynamic parameter, is a clean and useful means to tune the interatomic distance and consequently, can be used to engineer the electronic and, subsequently, the macroscopic physical properties of the system. In addition, it is possible to trigger novel structural and/or electronic transitions. Indeed, we recently observed pressure-induced topological phase transitions and even superconductivity in topological materials.[24-27] In this work we study the effect of pressure on the electrical transport properties and crystal structures of $MnBi_2Te_4$ and $MnBi_4Te_7$ in a diamond anvil cell (DAC) apparatus. The antiferromagnetic (AFM) metallic ground state of $MnBi_2Te_4$ and $MnBi_4Te_7$ single crystal is gradually suppressed by pressure, and the conductance as well as the crystal structure change dramatically upon further compression. Through *ab initio* band structure calculations, we found that the application of pressure does not qualitatively change the electronic and topological nature of the two systems until the structural phase transition observed in the high-pressure regime. Based on synchrotron XRD and Raman spectroscopy measurements, detailed high-pressure crystal structure and phase transitions are discussed.

The $MnBi_2Te_4$ and $MnBi_4Te_7$ single crystals in this work were grown using a flux-assisted method.[12] High pressure resistivity measurements were performed in a nonmagnetic DAC. A cubic BN/epoxy mixture layer was inserted between BeCu gaskets and electrical leads. Four Pt foils were arranged in a van der Pauw four-probe configuration to contact the sample in the chamber for resistivity measurements. NaCl was used as pressure transmitting medium (PTM) and pressure was determined by the ruby luminescence method.[28]

An in situ high pressure Raman spectroscopy investigation of $MnBi_2Te_4$ and $MnBi_4Te_7$ was performed using a Raman spectrometer (Renishaw inVia, U.K.) with a laser excitation wavelength of 532 nm and low-wavenumber filter. A symmetric DAC with anvil culet sizes of 400 μm was used, with silicon oil as the PTM. In situ high pressure XRD measurements were performed at beamline 13-BM-C of the Advanced Photon Source (APS) (X-ray wavelength $\lambda = 0.4340$ Å) and beamline BL15U of Shanghai Synchrotron Radiation Facility (X-ray wavelength $\lambda = 0.6199$ Å). Symmetric DACs with anvil culet sizes of 400 μm and 300 μm and T301 gaskets were used. Neon was used as the PTM and pressure was determined by the ruby luminescence method.[28] The two-dimensional diffraction images were integrated into angle-resolved diffraction intensity profiles using the software DIOPTAS.[29] Rietveld refinements on crystal structures under high pressure were performed using the General Structure Analysis System (GSAS) and the graphical user interface EXPGUI.[30]

The *ab initio* calculations were performed within the framework of density functional theory (DFT) as implemented in the Vienna *ab initio* simulation package (VASP),[31] with the exchange-correlation functional considered in the generalized gradient approximation potential.[32] A k-mesh of 9×9×1 for $MnBi_2Te_4$ and 9×9×3 for $MnBi_4Te_7$ was applied. The experimental lattice constants were adopted under different pressures with atomic positions optimized for a total energy-tolerance of $10^{-5}$ eV. To account for the correlation effect of the transition metal element Mn in both $MnBi_2Te_4$ and $MnBi_4Te_7$, the LDA+U functional with $U = 3$ eV for the *d*-orbitals of Mn is adopted. The spin-orbit coupling was considered self-consistently in this work. The topological surface states were calculated by applying the iterative Green's function approach[33] as implemented in WannierTools[34] based on the maximally localized Wannier functions[35] as obtained through the VASP2WANNIER90[36] interfaces in a non-self-consistent calculation.

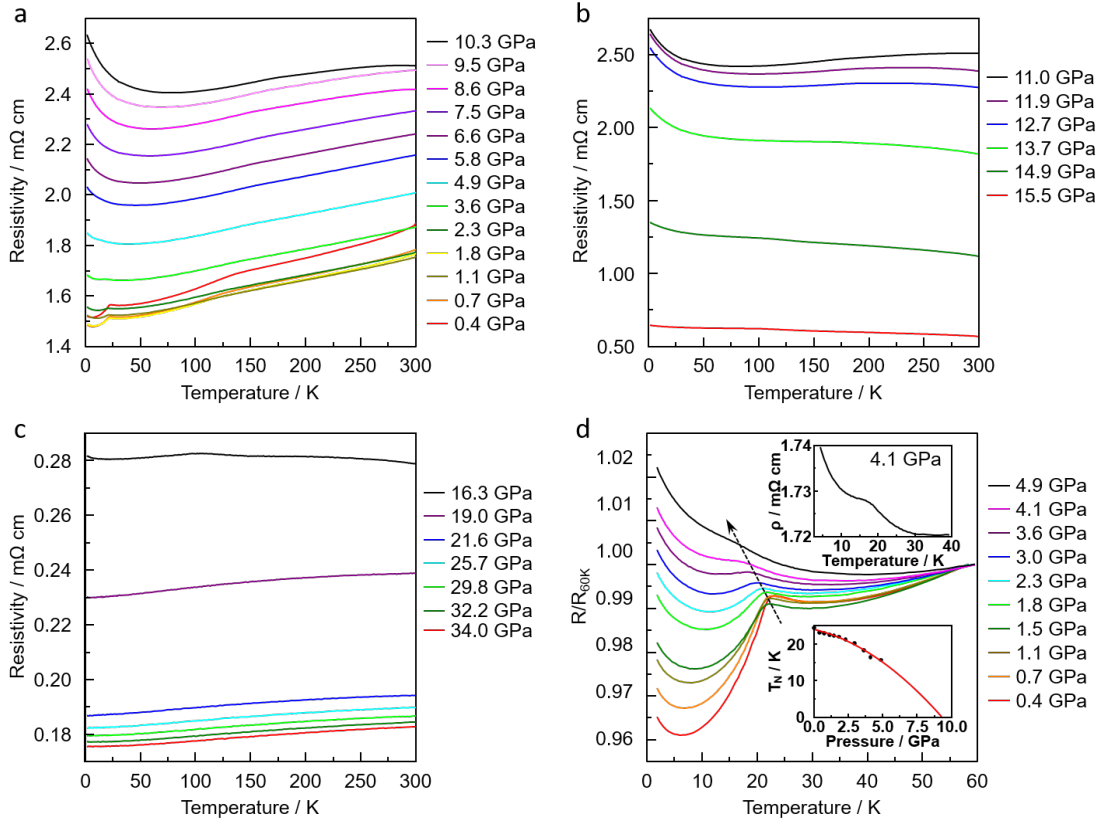

Fig. 2. Electrical resistivity of MnBi$_2$Te$_4$ as a function of temperature for pressures up to 10.3 GPa (a), 15.5 GPa (b) and 34.0 GPa (c); (d) Detail of the normalized resistivity of MnBi$_2$Te$_4$ as a function of temperature at various pressures to monitor the shift of the AFM transition kink. The inset shows the enlarged resistivity-temperature curve at 4.1 GPa and fitting of the AFM transition temperature as a function of pressure.

As a typical layered material, the electrical transport and magnetic properties of MnBi$_2$Te$_4$ and MnBi$_4$Te$_7$ are expected to be sensitive to the competition between interlayer and intralayer interactions, which can be effectively tuned by applying external pressure. We performed resistivity measurements on several single crystals at various pressures. Figure 2a, b, and c show the typical $\rho(T)$ curves of MnBi$_2$Te$_4$ for pressures up to 34.0 GPa. As shown in Figure 2a, the resistivity-temperature slope $dR/dT$ of MnBi$_2$Te$_4$ clearly shows a positive value, indicating metal-like conduction in the low-pressure range. With an increase of pressure, $\rho(T)$ curves show an upturn behavior at low temperatures. Upon further compression, a metal-semiconductor transition is observed and $\rho(T)$ displays a semiconductor-like behavior for $P > 12$ GPa. Interestingly, the resistivity ultimately undergoes a metallization at a pressure above 16.3 GPa and does not change significantly in response to further increases in the pressure. No superconductivity was observed down to 1.8 K in this pressure range.

It should be noted that $\rho(T)$ of MnBi$_2$Te$_4$ displays a kink at the AFM transition $T_N = 24.5$ K at 0.4 GPa (Figure 2d) consistent with the magnetic measurements shown in Figure S1 and those of other reports.[20] The rapid drop of resistivity below $T_N$ is attributed to the reduction of spin scattering after the formation of long-range AFM order.[37] As indicated by the arrow in Figure 2d, $T_N$ determined from the resistivity kink shifts to lower temperatures with increasing pressure. Over 4.9 GPa, the upturn resistivity trend at lower temperature becomes much stronger and the kink merges into the $\rho(T)$ curve. The fitting results demonstrate that $T_N$ approaches zero at approximately 9.3 GPa. Since the interlayer distance deceases under high pressure, it is speculated that the pressure-induced enhancement of antiferromagnetic/ferromagnetic competition and the partial delocalization of Mn-3$d$ electrons not only destroys long-range AFM order, but also promotes charge-carrier localization through enhanced spin fluctuations and/or the formation of a hybridization gap at high pressure.

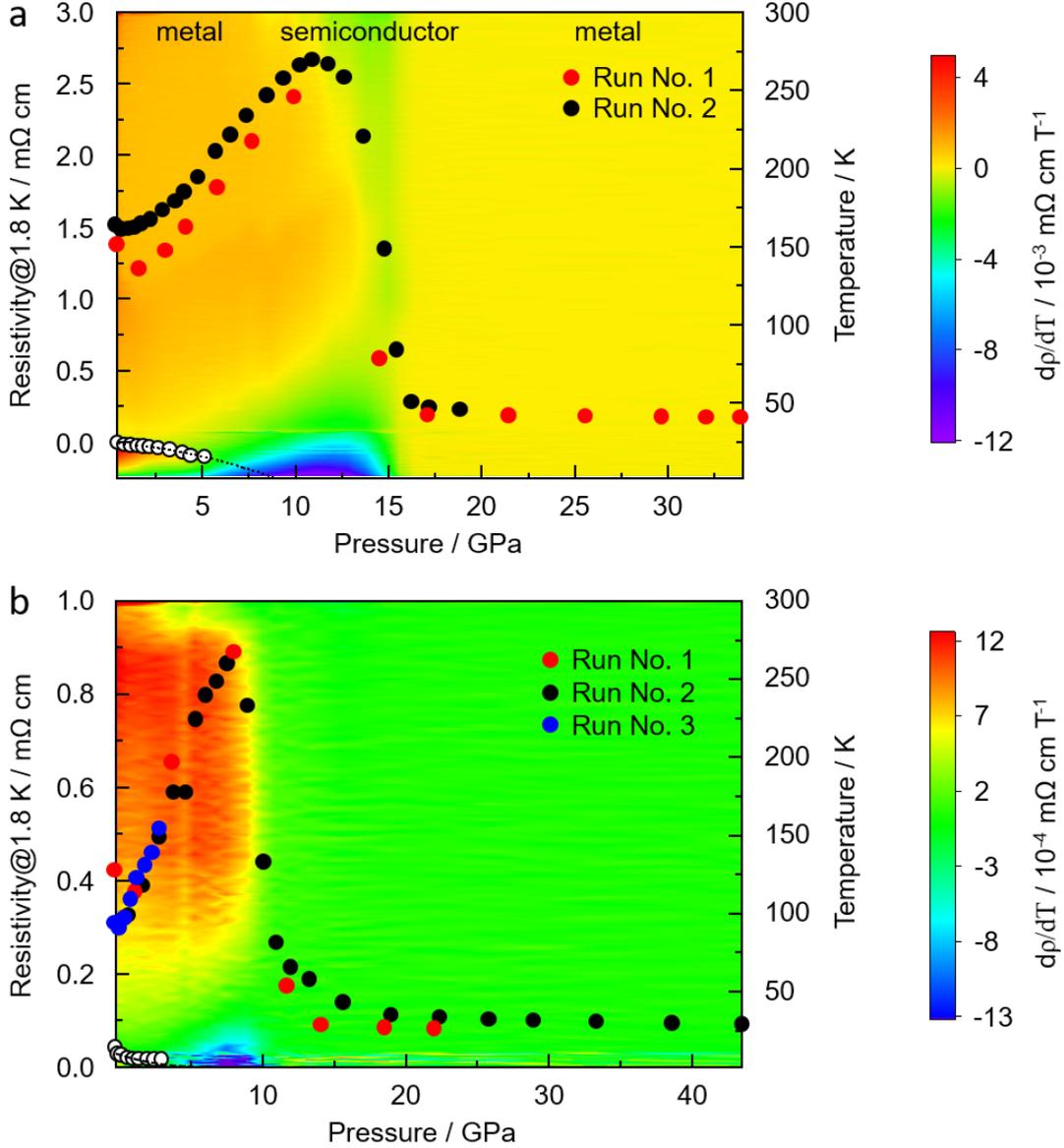

Fig. 3. Electronic phase diagram of MnBi$_2$Te$_4$ (a) and MnBi$_4$Te$_7$ (b), respectively. The black and red solid circles represent different runs of electrical resistivity measurements at 1.8 K. The black open circles indicate AFM transition temperatures due to the transport measurements.

The high-pressure experiments have been repeated on different samples with good reproducibility of the observed transition temperatures. Based on the above resistivity measurements, we summarize a *T-P* phase diagram for MnBi$_2$Te$_4$ single crystals in Figure 3a. The resistivity of MnBi$_2$Te$_4$ shows non-monotonic evolution with increasing pressure. Over the entire temperature range, the resistivity is first suppressed with applied pressure and reaches a minimum value at about 2 GPa. As the pressure further increases, the resistivity increases with a maximum occurring at 11.0 GPa and the AFM order shifted to a lower temperature. Accompanying the suppression of the AFM transition, the electrical transport properties also change qualitatively from metal-like *dρ/dT* > 0 to semimetal- or semiconducting-like behavior *dρ/dT* < 0. For *P* > 12 GPa, the resistivity abruptly decreases and a transition from semiconducting to metallic behavior takes place at further increased pressure. Similarly, pressure-induced non-monotonic evolution was also observed in MnBi$_4$Te$_7$, as shown in Figure 3b. Although resistivity changes significantly under high pressure, *ρ(T)* exhibits a metallic behavior over the whole temperature range (Figure S2). No transition from metallic to semiconducting behavior was observed within the studied pressure range. The AFM order of MnBi$_4$Te$_7$ shifted to a lower temperature with increase pressure, which was similar to that of MnBi$_2$Te$_4$ (Figure S3).

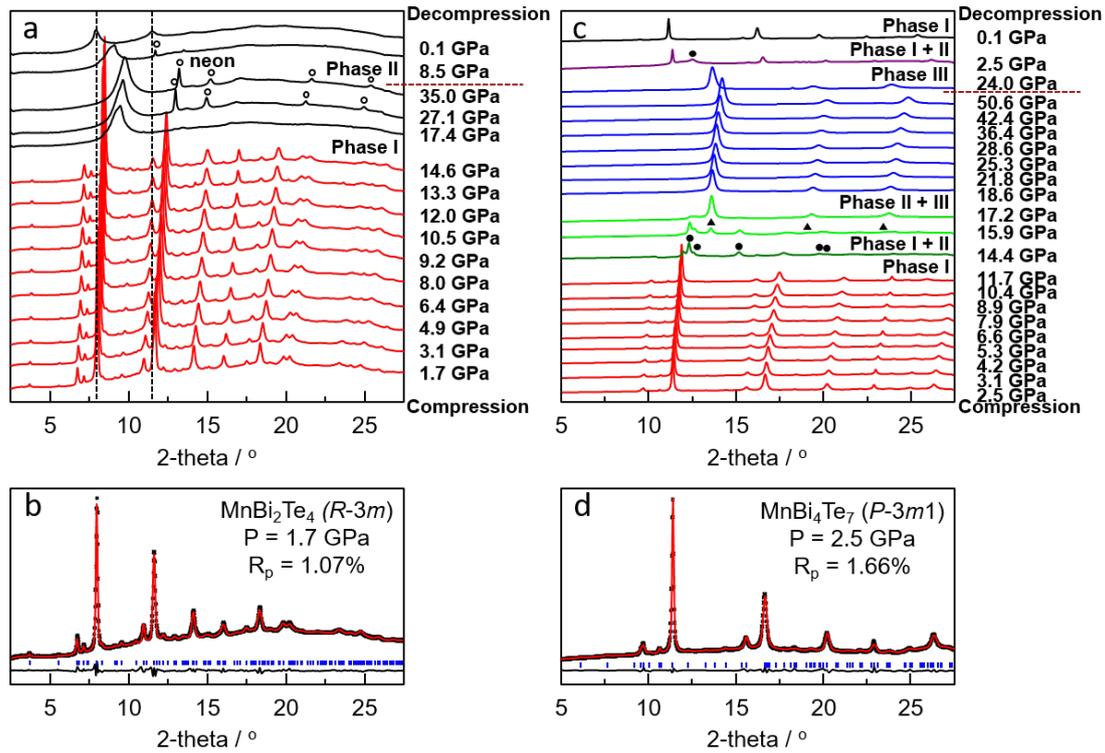

Fig. 4. XRD patterns collected at various pressure for MnBi$_2$Te$_4$ with an X-ray wavelength of $\lambda = 0.4340$ Å (a) and MnBi$_4$Te$_7$ with an X-ray wavelength of $\lambda = 0.6199$ Å (c); The black open circles are from PTM neon; Typical Rietveld refinement of phase I of MnBi$_2$Te$_4$ (b) and MnBi$_4$Te$_7$ (d), respectively. The experimental and simulated data are indicated by black stars and red lines, respectively. The solid lines shown at the bottom of the figures are the residual intensities. The vertical bars indicate peak positions.

The angular dispersive XRD patterns of MnBi$_2$Te$_4$ at various pressures are shown in Figure 4a. Under ambient conditions and in the low-pressure range ($P \leq 14.6$ GPa), all the diffraction peaks of MnBi$_2$Te$_4$ could be indexed to the rhombohedral *R-3m* (no. 166) structure by Rietveld refinement (Figure 4b). High pressure XRD experiments in pressure steps of 1-2 GPa were performed on MnBi$_2$Te$_4$ via a DAC. At pressures exceeding 14.6 GPa structural disorder becomes apparent. Above 17.4 GPa, diffraction peaks from crystalline phase disappear, and a new broad peak appears at approximately 2.65 Å in *d*-spacing. This indicates that the sample has completely transformed into an amorphous state.

In contrast, a different structure evolution for MnBi$_4$Te$_7$ is observed under high pressure (Figure 4c). In the low-pressure range, phase I of MnBi$_4$Te$_7$ crystallizes in a trigonal space group *P-3m*1 (no. 164) as shown in Figure 4d. At 14.4 GPa, a high-pressure (HP) phase, phase II, was observed. This phase is only stable in a narrow pressure range and coexists with the phase I or the phase III upon compression. Above 18.6 GPa, only phase III exists and no further transitions are observed up to 50.6 GPa. Upon decompression, phase III persists to 24.0 GPa. When the pressure is decreased to 2.5 GPa, phase II and phase I are recovered and coexist. After a full pressure release, MnBi$_4$Te$_7$ recovers the ambient-pressure structure.

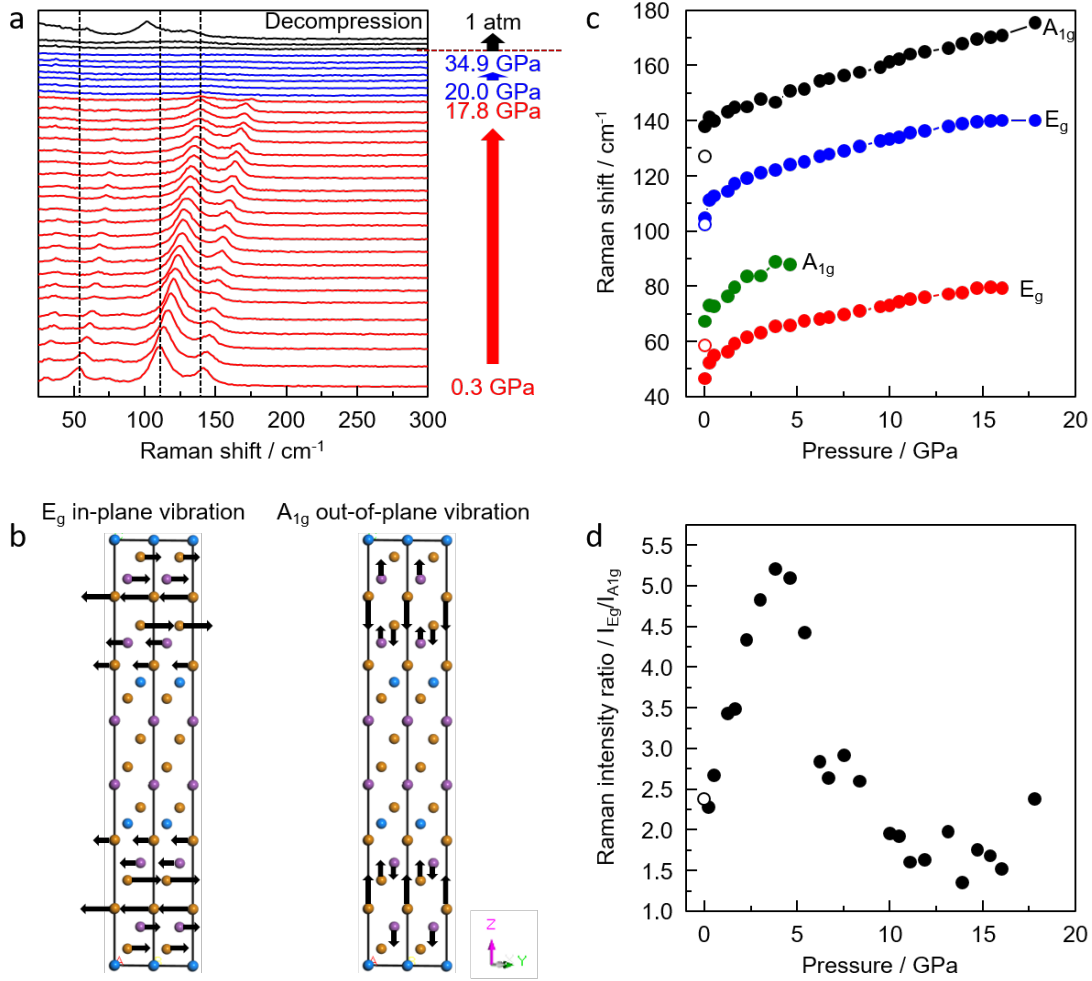

Fig. 5. (a) Raman spectra at various pressures for MnBi$_2$Te$_4$; (b) Phonon mode symmetry and direction of vibration for MnBi$_2$Te$_4$; (c) Raman mode frequencies for MnBi$_2$Te$_4$ in compression (solid circle) and decompression (open circle); (d) Pressure dependence of $I_{Eg}/I_{A1g}$ intensity ratio for MnBi$_2$Te$_4$. For accurate peak intensity comparison, the strong $E_g$ and $A_{1g}$ modes with respective Raman shifts of 104.2 cm$^{-1}$ and 139.8 cm$^{-1}$ at 1 atm are chosen. Peak intensity and peak position are obtained by Gaussian and Lorentzian mixed line shape fitting. Open circles represent data recovered for 1 atm.

To verify our speculation on the crystallographic structural phase transition sequence under high pressure, Raman scattering spectroscopy was employed to characterize the pressure-induced phase transition (Figure 5a). According to group theory analysis and the results in the literature,[38] there are four Raman-active modes ($2E_g + 2A_{1g}$) for MnBi$_2$Te$_4$. The $E_g$ and $A_{1g}$ modes are related to the in-plane A(VI)-B(V) and out-of-plane lattice vibrations, respectively (Figure 5b). At 0.3 GPa, four peaks are assigned as follows: 47.4 cm$^{-1}$ ($E_g$), 67.4 cm$^{-1}$ ($A_{1g}$), 104.2 cm$^{-1}$ ($E_g$), and 139.8 cm$^{-1}$ ($A_{1g}$).[38] As the pressure is increased, all four modes exhibit blue-shift due to the increase in the strength of the Bi-Te covalent interaction (Figure 5c). Upon further compression exceeding a pressure of 17.8 GPa, all the peaks disappeared. The pressure-induced amorphization occurs at 17.8 GPa, which coincides with the XRD result at 17.4 GPa. In addition, a reversible phase transition associated with a compressed lattice (where the lattice constants are decreased) is verified by the Raman spectrum of the sample after recovery to 1 atm. The Raman spectra of MnBi$_4$Te$_7$ were also measured using a DAC and a similar phenomenon was observed under high pressure (Figure S4). It should be noted that no new Raman modes were observed under higher pressure, although a structural phase transition is observed by synchrotron XRD measurements. One can expect that pressure-induced metallization or vibration modes become weaker under high pressure, which may account for the absence of Raman modes.

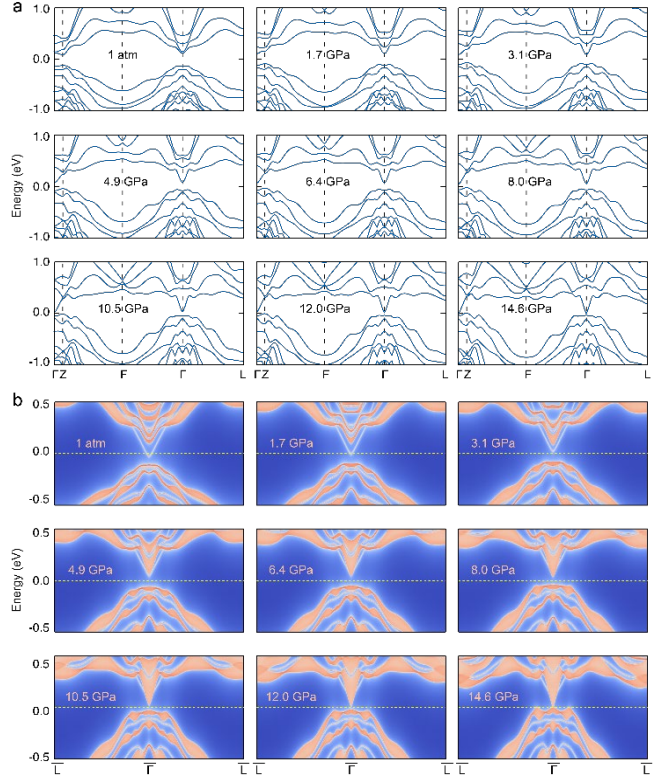

Fig. 6. Bulk and surface electronic structures of MnBi$_2$Te$_4$ at different pressures. (a) The bulk electronic structure remains gapped under all applied pressures with roughly monotonic decrease of gap size. (b) The topological surface states on (001) display a reentrant behavior upon the increase of pressure.

To understand the non-monotonic change of the measured resistivity under different pressures, we performed detailed *ab initio* calculations and examined both the bulk and surface electronic structure of MnBi$_2$Te$_4$ (Figure 6 and S5). Electron transportation is mainly determined by the states around the Fermi level which can be effectively tuned by external pressure. Concerning a topological system, these states contain both the bulk and the topological surface/edge contributions. It is widely known that the topology of a topological system is fully determined by the symmetry and the associated Berry curvature of the bulk bands. As long as they are qualitatively unchanged, the topology persists (Figure S5). However, external perturbations, such as the pressure, can modify the dispersions of both the bulk and surface bands, resulting in different transport responses. In Figure 6a the electronic structures of MnBi$_2$Te$_4$ are displayed for different pressures. MnBi$_2$Te$_4$ is a semiconductor with a gap of 243.4 meV at atmospheric pressure. Once external pressure is applied, the conduction band bottom changes from $Z$ to $\Gamma$ and the gap size gradually decreases with increasing pressure. At the highest pressure applied which maintains the crystal symmetry of MnBi$_2$Te$_4$, a global gap remains but is significantly reduced to 16.3 meV.

Due to the topological nature of this system, the total conductance/resistance experimentally measured is subjected to contributions from both the bulk and surface electrons. We, thus, further determined the surface electronic structure of the experimentally cleaved (001) surface (Figure 6b). In sharp contrast to the bulk electronic structure, an overall reentrant behavior of the gapped surface states is observed upon the increase of pressure. Below 4.9 GPa, the magnetic surface states move from below to above the Fermi level, leading to a metal-semiconductor transition solely caused by the surface electrons. The surface states, with a clear separation from the bulk bands at pressures below 4.9 GPa, completely merge into the bulk bands at pressures above 10.5 GPa. Above this pressure, the contribution to the resistivity is mainly determined by the bulk gap and electrons. Thus, the decrease of the bulk gap results in a decrease of the resistivity as shown in Figure 3a above 10.5 GPa. While, below 10.5 GPa, the bulk and surface electrons behave differently, i.e. the surface electrons are gradually localized, while the bulk electrons become more mobile with increasing pressure. Thus, the competition between the two types of electrons results in the decrease-increase behavior of the resistivity observed in Figure 3a. More precisely, we suspect that the decrease of the resistivity below 3.1 GPa is mainly induced by the delocalization of the bulk electrons as the surface electrons remain metallic; while the increase of the resistivity is mainly a consequence of the localization of the surface electrons, as the surface states are no longer metallic and have not yet merged into the bulk states.

    A similar analysis can be applied to $MnBi_4Te_7$ (Figure S5 and S6). We note that the bulk gap shown in Figure S6a decreases under increasing pressure with the minimum gap decreasing from 196.7 meV to 174.8 meV at $\Gamma$, which only renders the more substantial contribution to its conductance at higher pressures. At low pressure, the contribution from the surface states becomes dominant. As $MnBi_4Te_7$ can be naturally cleaved at $MnBi_2Te_4$ and $Bi_2Te_3$ layers, the total surface conductance includes components from both terminations. The topological surface states with $MnBi_2Te_4$ termination intersect the Fermi level under all examined pressures (Figure S6b), presenting a metallic background in the measured pressure range. Meanwhile, the surface electrons terminated at $Bi_2Te_3$ become more localized with increasing pressure. The surface band crosses the Fermi level at 2.5 GPa, and gradually moves to higher binding energies with further increase of pressure. The competition between the two types of electrons again results in the decrease-increase behavior of resistivity observed in Figure 3b. Below 2.5 GPa, the delocalization of the bulk electrons is attributed to the decrease of resistivity, while the subsequent increase of the resistivity above 2.5 GPa mainly stems from the localization of the surface electrons. At approximately 10.4 GPa, the surface bands merge into the bulk states, after which the resistivity is mainly determined by the bulk electrons, and thus shows a sharp decline as observed in the transport measurements.

    Next, we discuss the high-pressure phase transition. To gain an insight into the structural evolution, pressure-induced bond length and angle variations in $MnTe_6$ and $BiTe_6$ octahedra are derived from the Rietveld refinements (Figure S7). The Mn-Te bond length decreases with an increase of pressure and the bond angle of Te(2)-Mn(1)-Te(2) changes only slightly (Figure S7a, S6b). Below 3.1 GPa, the pressure coefficients of Bi(1)-Te(1) and Bi(1)-Te(2) bond lengths show the same sign. Also, both the Te(1)-Bi(1)-Te(1) bond angle and the distance between the two Te(1)s of the near neighbor septuple blocks decrease with increasing pressure (Figure S7c, S7d). However, Bi(1)-Te(2) and Bi(1)-Te(1) bond lengths show opposite responses to pressure near 4 GPa, and Te(1)-Bi(1)-Te(1) bond angle also behaves differently from Te(1)-Bi(1)-Te(2) and Te(2)-Bi(1)-Te(2) bond angles (Figure S7c, S7d). These results indicate that the distorted octahedral $BiTe_6$ deforms more significantly above 3.1 GPa and bends towards Mn atoms (Figure S7e). This releases stress along the *c*-axis and weakens the interaction between nearest neighbor SL. Upon further compression, interlayer interaction enhancement dominates and ultimately the interlayer Te(1)-Te(1) bond is formed. The pressure dependence of the *c*/*a* ratio for the *R*-3*m* phase of $MnBi_2Te_4$ (plotted in Figure S8a) shows a minimum at approximately 3.1 GPa. This is consistent with the Raman spectroscopy observations. The in-plane Bi-Te vibration was enhanced significantly with pressure. With further compression, the ratio of out-of-plane Bi-Te vibrations is enhanced when the pressure exceeds 3.8 GPa as shown in the evolution of the intensity ratio ($I_{Eg}/I_{A1g}$) in Figure 5d. A similar phenomenon was observed in the $MnBi_4Te_7$ phase (Figure S4d). It is clear that the structural change induced by external pressure will significantly modify the corresponding electronic structure. The reduced interlayer distance will enhance the three-dimensional dispersion of the system. As a result, the electronic states around the Fermi level will be considerably modified, which, ultimately, influences the macroscopic resistivity. Thus, such layered MTIs with large inter-layer distances are ideal for pressure engineered materials.

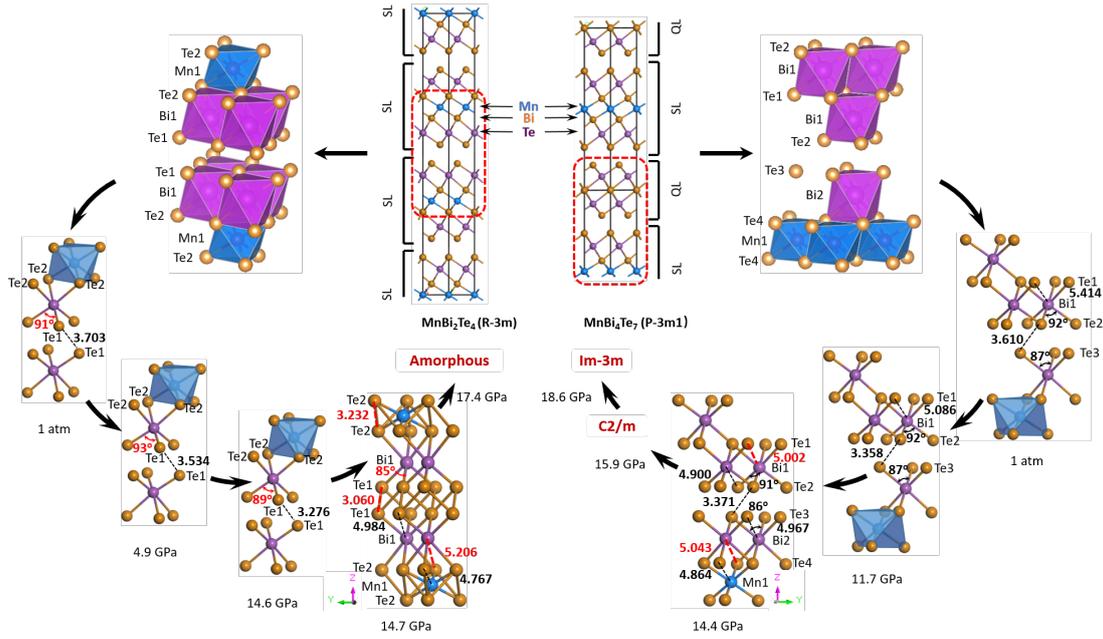

Fig. 7. Schematic representation of the evolution of $MnBi_2Te_4$ and $MnBi_4Te_7$ structures under high pressure.

In compression, there is quite a distinct compression behavior between $MnBi_2Te_4$ and $MnBi_4Te_7$. Figure 7 shows the pressure-induced structural evolution of $MnBi_2Te_4$ and $MnBi_4Te_7$, respectively. The former transforms to an amorphous phase at approximately 17.4 GPa, while the latter transforms from a rhombohedral to a mixed HP phase at 14.4 GPa, and finally phase III obtained at 18.6-50.6 GPa. For phase II of $MnBi_4Te_7$, Le Bail refinement yielded a monoclinic structure with $a$ = 14.4192(3) Å, $b$ = 3.9415(9) Å, $c$ = 17.12027(9) Å, $\beta$ = 148.62(7) °. The XRD pattern of phase III is simple and can be indexed to a $Im$-$3m$ (no. 229) structure with $a$ = 3.6796(0) Å (Figure S9, Table S1). The different compression behavior is related to the distortion of $MnTe_6$ and $BiTe_6$ which is induced by competition under high pressure. In $MnBi_2Te_4$, the Te(2)-Te(2) bond forms besides Te(1)-Te(1) linked in the low-pressure range accompanied by $MnTe_6$ octahedron flattening. In contrast, the distance of Bi(1)-Te(1) (5.002 Å) and Bi(1)-Te(2) (4.900 Å) is shorter than Bi(2)-Te(4) (5.043 Å) in $MnBi_4Te_7$ and Bi(1)-Te(2) (5.206 Å) in $MnBi_2Te_4$. As a result, the pressure-induced distorted Bi(1)Te$_6$ octahedron in the $Bi_2Te_3$ quintuple block tends to form a heptahedrally-coordinated $BiTe_7$ unit and further Bi(Te)$_n$ (n > 7). At 18.6 GPa, the Bi-Bi and Bi-Te distance in $MnBi_4Te_7$ are close to each other because of the flatter $MnTe_6$ octahedron as well as the improved interaction between QL and SL (Figure S10). An alternating Bi, Te structure with Mn intercalation exists during the formation of an isotropic phase along the layers and perpendicular to the layers. The structural evolution of $MnBi_4Te_7$ under high pressure resembles the situation in the case of $Bi_2Te_3$[8, 39, 40]. Recent sister compounds $MnBi_6Te_{10}$ ($m$ = 1, $n$ = 2) and $MnBi_8Te_{13}$ ($m$ = 1, $n$ = 3) have been grown successfully.[41-44] It is will be interesting to characterize the structural evolution of this series $(MnBi_2Te_4)_m(Bi_2Te_3)_n$ of compounds and summarize pressure-induced phase transition in these layered compounds.

In conclusion, we have performed a comprehensive high-pressure study on the electrical transport properties and crystal structures of the MTIs $MnBi_2Te_4$ and $MnBi_4Te_7$ in DACs. The AFM metallic ground state of $MnBi_2Te_4$ and $MnBi_4Te_7$ single crystals are gradually suppressed by pressure. The pressure-dependent resistivity over a wide temperature range passes through a minimum at around 3 GPa. Upon further increasing the pressure, resistivity starts to increase rapidly, reaching a maximum at a pressure above 10 GPa. Through *ab initio* calculations, we find that the application of pressure does not destroy the nontrivial topology of the system before structural phase transition. However, the bulk and surface states respond differently to external pressure, resulting in competing contributions to the macroscopic resistivity. Based on synchrotron XRD and Raman spectroscopy measurements, we found that $MnBi_2Te_4$ transforms to an amorphous phase at around 17.4 GPa, while MnBi$_4$Te$_7$ transforms to two new high-pressure phases. Application of pressure effectively tuned the electronic properties and crystal structure of MnBi$_2$Te$_4$ and MnBi$_4$Te$_7$. Considering both intriguing magnetism and topology in this layered material, our results call for further experimental and theoretical studies on (MnBi$_2$Te$_4$)$_m$(Bi$_2$Te$_3$)$_n$ and related materials for a better understanding of the interplay between magnetic and topological nature, and its potential application in realizing topological superconductivity.

[*]Supported by the National Key Research and Development Program of China under Grant Nos. 2018YFA0704300 and 2017YFE0131300, the National Science Foundation of China under Grant Nos. U1932217, 11974246 and 11874263 and Natural Science Foundation of Shanghai under Grant Nos. 19ZR1477300. The authors thank the support from Analytical Instrumentation Center (# SPST-AIC10112914), SPST, ShanghaiTech University. This work was partially supported by Collaborative Research Project of Materials and Structures Laboratory, Tokyo Institute of Technology, Japan. Part of this research is supported by COMPRES (NSF Cooperative Agreement EAR-1661511). We thank Dr. Zhenhai Yu for valuable discussions. We thank the staffs from BL15U1 at Shanghai Synchrotron Radiation Facility, for assistance during data collection. Portions of this work were performed at GeoSoilEnviroCARS (The University of Chicago, Sector 13), Advanced Photon Source (APS), Argonne National Laboratory. GeoSoilEnviroCARS is supported by the National Science Foundation-Earth Sciences (EAR-1634415) and Department of Energy-GeoSciences (DE-FG02-94ER14466). This research used resources of the Advanced Photon Source, a U.S. Department of Energy (DOE) Office of Science User Facility operated for the DOE Office of Science by Argonne National Laboratory under Contract Nos. DE-AC02-06CH11357.National Natural Science Foundation of China under Grant Nos. 10225417 and the National Basic Research Program of China under Grant Nos. 2006CB601003.